# FILTERGRAPH: AN INTERACTIVE WEB APPLICATION FOR VISUALIZATION OF ASTRONOMY DATASETS

Dan Burger[1], Keivan G. Stassun[1,2], Joshua Pepper[1,3], Robert J. Siverd[1], Martin Paegert[1], Nathan M. De Lee[1], William H. Robinson[4]

[1]Department of Physics & Astronomy, Vanderbilt University, VU Station B 1807, Nashville, TN 37235

[2]Department of Physics, Fisk University, 1000 17th Ave. N., Nashville, TN 37208

[3]Department of Physics, Lehigh University, 27 Memorial Drive West, Bethlehem, PA 18015

[4]Department of Electrical Engineering and Computer Science, Vanderbilt University, 2301 Vanderbilt Place PMB 351826, Nashville, TN 37235

**Abstract:** Filtergraph is a web application being developed and maintained by the Vanderbilt Initiative in Data-intensive Astrophysics (VIDA) to flexibly and rapidly visualize a large variety of astronomy datasets of various formats and sizes. The user loads a flat-file dataset into Filtergraph which automatically generates an interactive data portal that can be easily shared with others. From this portal, the user can immediately generate scatter plots of up to five dimensions as well as histograms and tables based on the dataset. Key features of the portal include intuitive controls with auto-completed variable names, the ability to filter the data in real time through user-specified criteria, the ability to select data by dragging on the screen, and the ability to perform arithmetic operations on the data in real time. To enable seamless data visualization and exploration, changes are quickly rendered on screen and visualizations can be exported as high quality graphics files. The application is optimized for speed in the context of large datasets: for instance, a plot generated from a stellar database of 3.1 million entries renders in less than 2 seconds on a standard web server platform. This web application has been created using the Web2py web framework based on the Python programming language. Filtergraph is free to use at http://filtergraph.vanderbilt.edu/.

**Highlights:**

- We developed a web-based application for visualization of astronomy data.
- The user can generate publication quality multi-dimensional plots and tables.
- Designed for speed, the user can instantly interact with and visualize the data.
- To enhance collaboration, data and visualizations can be shared via simple URL.
- This web application can accept data in a wide variety of file formats.

1. Introduction

Increasingly in astronomy there is a need for performing quick-look inspection and visualization of large datasets in order to easily ascertain the nature and content of the data, begin identifying possibly meaningful structures or patterns in the data, and guide more computationally costly deep-dive analyses of the data. For example, consider that the first product of a large survey project is often a large database with many columns (representing the various measurable and/or derived quantities) and with many rows (representing the individual objects of study); it is not uncommon for such databases to include tens of columns and millions of rows.

To even begin visualizing such datasets---let alone perform basic analyses---can be a daunting task. The researcher is faced with questions such as: What is the content and what does it look like? Where are the "holes" in the data (missing or bad data) and are there systematics or biases to be wary of? What are the relationships among the variables in the dataset, and are they meaningful? Are there interesting patterns that might be worthy of deeper investigation and analysis?

Indeed, the large size and high dimensionality of such datasets make visualization challenging. Identifying potentially meaningful patterns often requires "seeing" the data simultaneously across multiple dimensions and with appropriate "slices" through several multidimensional spaces.

Arguably even more fundamental to the visualization challenge is what might be called the high "potential barrier" that the user faces to even begin the visualization process. Certainly there exist high-end tools for storing data in a database, plotting data, and so on. But for many researchers there is a very large overhead associated with using such tools from the first instance: importing the data and correctly specifying meta-data, keeping track of a large number of variable names, issuing and scripting commands for plotting, plotting pairwise variables against one another over multiple iterations, attempting to filter out bad data, re-rendering plots to restricted data ranges, attempting to represent many variables at once, and so on. Faced with such high overheads in time and effort, there is the temptation to either skip the crucial quick-look visualization step altogether, or else to make very limited attempts at representing the data with simple plots based on preconceptions about what should be meaningful to visualize.

We have developed Filtergraph as "Plotting 2.0", an easy-to-use web-based solution to this problem. The principal motivation for Filtergraph is to reduce the "activation energy" of the data visualization process. Users can register to use the tool instantly, can immediately upload datasets without the requirement of meta-data specification, and can thus begin seeing their data in seconds. We have also sought to make Filtergraph intuitive (see the sample in Figure 1). Plots involving two, three, four, or five dimensions (3D + color + symbol size), as well as histograms, can be generated with a few clicks, and variable name fields are pre-filled and auto-complete so that the user does not need to remember the full content of the dataset in order to make plots. Mathematical operations on individual variables---or indeed among variables---can be performed on the fly. For instance, if there exist two columns in a dataset labeled "var1" and "var2", one can simply specify the x-axis of a plot as "var1 – var2". In addition, data can be easily filtered by specifying data ranges explicitly or by dragging over a region of interest on the screen.

Importantly, Filtergraph is fast. There is nothing more frustrating or deterrent to the creative visualization process than being faced with long lag times between subsequent plot renderings. If plots do not update instantly, users will be more likely to avoid the penalty associated with trial and error exploration. Under optimal conditions, Filtergraph renders a plot of 1.5 million points in approximately one second, enabling and encouraging natural, seamless interaction with and exploration of the data. The user can change variables, attempt different mathematical operations, filter in and out, move back and forth between different representational forms, over and over, in seconds without cognitive interruption.

Finally, Filtergraph is designed to facilitate sharing. Any Filtergraph plot can be saved as a graphics file in various formats. Filtered subsets of the data can also be saved as tables in various formats. More importantly, each dataset is instantly set up as a sharable "data portal" (http://filtergraph.vanderbilt.edu/portal_name) that can be provided to collaborators. Instead of sending collaborators a copy of the raw data file, the user can easily provide a simple URL that contains the data and the ability to instantly visualize it, thus greatly facilitating the collaboration process. With few exceptions, portals can be accessed by anyone who has the URL.

There are data visualization services with similar interfaces developed for specific purposes, such as the Exoplanet Data Explorer [1] [2] and Gapminder [3]. However, the data on these services are fixed. Filtergraph allows the user to visualize any given set of data. In addition, an advantage of Filtergraph over desktop plotting software is outsourcing the plotting operations to an external server and allowing the resulting plot to be instantly shared.

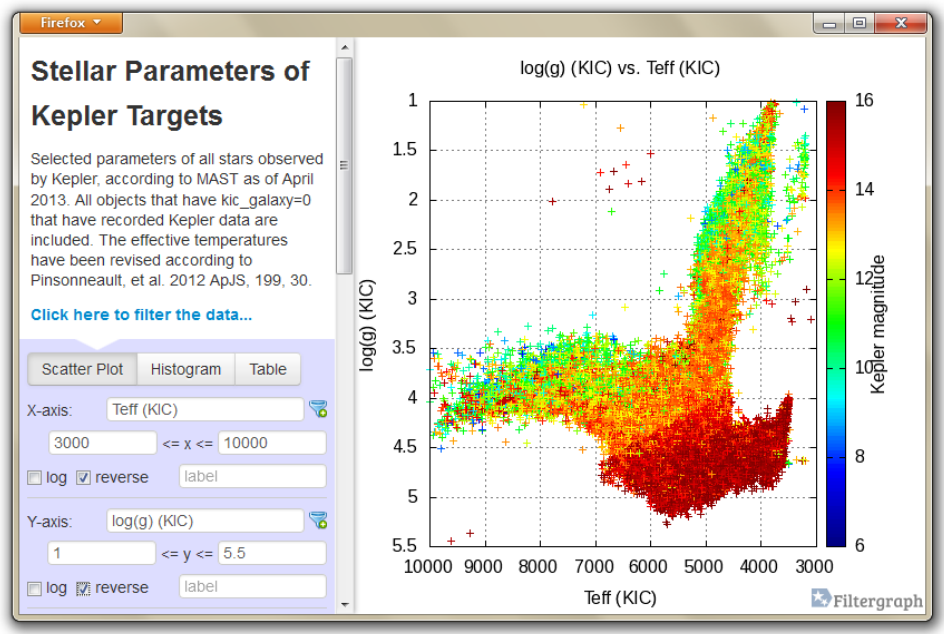

Figure 1. A screenshot of a Filtergraph portal showing properties of stars observed by the Kepler mission [4] [5]. This portal is publicly available at http://filtergraph.vanderbilt.edu/kepstars/. A version of this portal with the settings listed above is available at http://filtergraph.vanderbilt.edu/2802807.

2. Current capabilities and use

Filtergraph is a free web-based service. On uploading a dataset to Filtergraph, the user is presented with a portal that can be used to generate plots and tables based on the dataset. The portal can be shared with others and provides interactive features such as zooming and obtaining information about a point on the scatter plot.

Filtergraph currently supports three visualization types: scatter plots, histograms and tables. A scatter plot consists of at least two columns: one for the X-axis and one for the Y-axis (see Figure 1). Three optional axes may be set. The color axis sets the color of each point on the basis of a range of colors; currently higher values are colored red, lower values are colored blue, and values in between are placed along the color spectrum. Similarly, the size axis sets the size of each point with higher values receiving larger points and lower values receiving smaller points. Selecting a Z-axis quantity transforms the scatter plot into a three-dimensional plot which can be viewed from different angles.

In cases where two or more points can overlap each other, a scatter plot may not be enough for adequately visualizing the data. Therefore, Filtergraph also provides the option of plotting the data as a one-dimensional histogram or two-dimensional heat map, each of which can be split on the basis of a set number of bins. The two-dimensional heat map can also be plotted as a three-dimensional surface with the ability to view from different angles. In addition, Filtergraph can also return the data as a table based on a selected subset of rows and columns in the dataset. This table can be sorted by a column in the table, ascending or descending. The output table can be easily shared with others using ASCII and HTML formats.

Filtergraph currently accepts uploads in several formats. Plain text ASCII files are accepted in space separated, comma separated (CSV), tab separated (TSV), and fixed width formats. In these cases, Filtergraph will automatically read the file to determine its properties: the file type, where the data begins in the file, the existence of a header line, the number of columns, and the format types for each column. Header names may be included but are not required. Filtergraph also accepts Microsoft Excel, SQLite, VOTable, FITS, IPAC, and Numpy file types. Upon upload, Filtergraph determines the structure of the dataset and populates all subsequent interfaces with the variable names determined from the header row, if provided.

      A Filtergraph portal consists of two parts, with the left sidebar being used to control the main content (see Figure 1). Key components of the left sidebar include: the name and description of the portal; the ability to switch between datasets on the portal, if more than one dataset is available; the ability to apply criteria, or filters, to the dataset; the ability to switch between the scatter plot, histogram, and table modes; the ability to apply display settings; the ability to output the data to a file (PNG, JPEG, GIF, Postscript, and PDF for graphs, HTML and ASCII for table data); and additional instructions and status information for using the dataset. User preferences and specific visualizations can also be saved to a URL for sharing and instant recall (the URL is auto-generated by Filtergraph).

      Axes are changed using an editable drop-down box that can include the name of a database column or mathematical functions on one or more database columns. The following functions are supported: addition, subtraction, multiplication, division, modulo, power, natural logarithm, base 10 logarithm, absolute value, square root, exponential, sine, cosine, tangent, and the hyperbolic and inverse versions of sine, cosine, and tangent. The constant pi (3.14159...) is also supported. Each axis is paired with an icon that allows the user to set its data range.

      Scatter plots and histograms are dynamic. Clicking on a point on the graph displays a pop-up window with all of the data for that particular point. Additionally, clicking and dragging on the window allows the user to zoom in on a particular region of the graph. A link is then provided to reset the zoom feature. Also, when the Z-axis is enabled for scatter plots, the user can rotate the 3D scatter plot.

2.1 Administrative interface

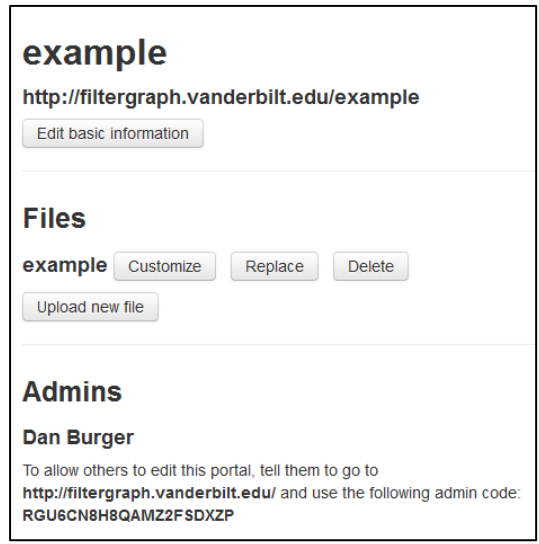

Figure 2. A screenshot of the administrative interface.

Filtergraph also provides a user-specific administrative interface (see Figure 2) for maintaining multiple portals, each of which may contain one or more datasets. An administrator for a portal can change or delete the portal and add, change or delete datasets associated with the portal. An administrator can also select another user to co-administer the portal by sharing a random string of characters (an "admin code") that is associated with each portal.

Once a user registers for the site, he or she is asked to create a portal. To create a portal, the user chooses a name and URL and then provides an initial dataset either by uploading a file or entering the URL of the data source. The current implementation limits users to 50MB per file; additional space may be provided upon request. More importantly, the software is not intrinsically limited by file size, and other choices of file size limits could be adopted.

Uploaded datasets are then inspected to determine the file type. For some file types such as ASCII, Filtergraph performs additional operations to determine the data types for each column. At this point, the portal is ready for use; the user may set the default appearance of the dataset as well as alternate names for each of its columns. Filtergraph also provides many standard user administration features provided by the Web2py web framework, such as changing profile information and obtaining lost usernames and passwords.

2.2 Customization

While Filtergraph provides a standard feature set, customizations are available upon request. For instance, the SLoWPoKES portal (http://filtergraph.vanderbilt.edu/slowpokes) [6] [7] [8] [9] [10] displays information from the Sloan Digital Sky Survey (SDSS) [11] gri composite image and the Two-Micron All Star Survey (2MASS) [12] H-band by clicking on a point on the graph. This is accomplished by interpreting XML data from the NASA/IPAC Infrared Science Archive (http://irsa.ipac.caltech.edu) and the SDSS-3 data site (http://skyserver.sdss3.org/dr8). In the past, one portal was configured to execute IDL scripts externally based on information about a given point.

3. Case studies

The original motivation for developing Filtergraph was to manage data coming from the version of the Kilodegree Extremely Little Telescope [13] located in Sutherland, South Africa (KELT-South) [14], which is a fully robotic telescope operated by Vanderbilt University and the South African Astronomical Observatory that searches for transiting exoplanets. KELT-South generates tens of thousands of images, each containing tens of thousands of stars, and analyzing even one of these images takes a significant amount of time. By using the web portal for KELT-South, we are able to select images for analysis effectively.

Filtergraph is finding broad use across many astronomical data visualization needs. Here we use the Hipparcos dataset [15] to illustrate a few representative case studies. A sample portal for Hipparcos has been set up at http://filtergraph.vanderbilt.edu/hiptest. As a first example, we generate a Hertzsprung-Russell diagram (see Figure 3).

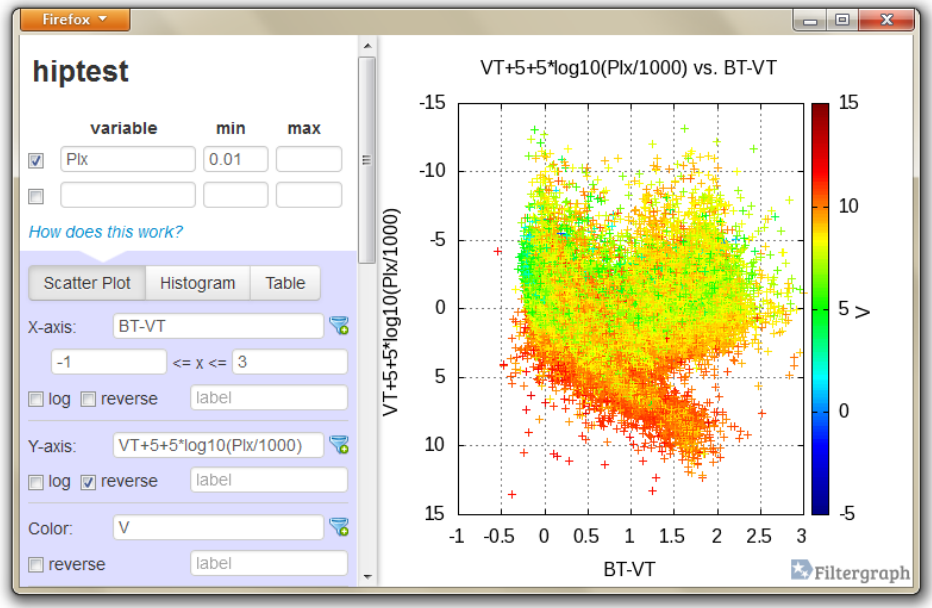

Figure 3. By applying the settings on the left to the publicly available Filtergraph portal at http://filtergraph.vanderbilt.edu/hiptest, the Hertzsprung-Russell diagram on the right can be generated. A version of this portal with these settings is available at http://filtergraph.vanderbilt.edu/3403816.

First, we filter out all points where the parallax variable (Plx) is less than 0.01. This is necessary so that all parallax data can be calculated using the logarithmic function presented in the Y-axis. For clarity, any outlier points where BT-VT is less than -1 or greater than 3 are not displayed, where BT and VT are the B and V magnitudes from the Tycho-2 catalog [16], respectively. The remaining points are plotted on the basis of the functions given for the X axis, Y axis, and color axis. The separation between dwarf stars and giant stars is apparent.

Beyond two-dimensional scatter plots, Filtergraph can produce a much wider variety of images. As an example, Figure 4 depicts a two-dimensional histogram produced using Hipparcos data [15], with colors assigned to each square region of the image on the basis of the density of data points in that region. In Figure 5, we use KELT-South data to generate a three-dimensional scatter plot with color as an additional axis. In the web interface, this three-dimensional display can be viewed from different angles using links at the bottom of the page.

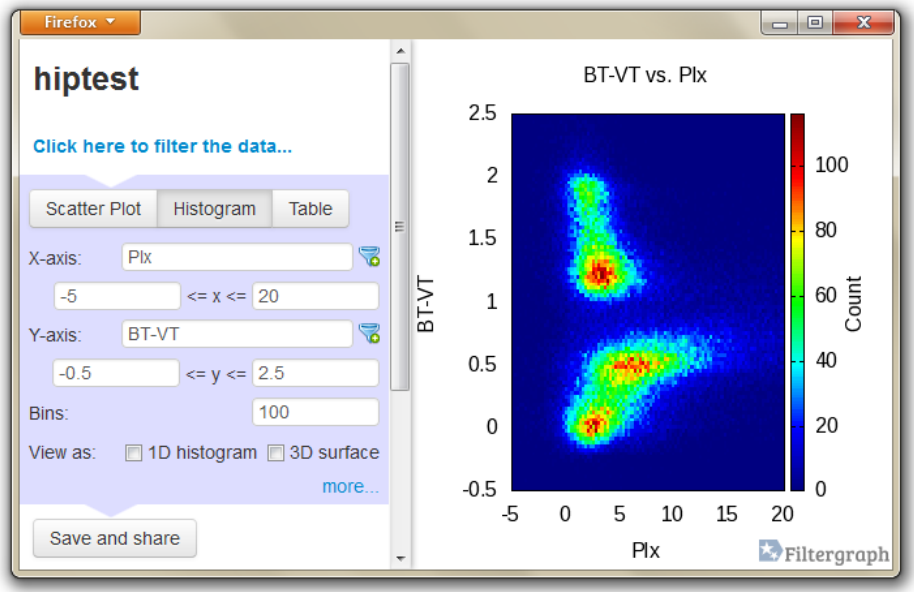

Figure 4. By applying the settings on the left to the publicly available Filtergraph portal at http://filtergraph.vanderbilt.edu/hiptest, the two-dimensional histogram on the right can be generated. A version of this portal with these settings is available at http://filtergraph.vanderbilt.edu/8789043.

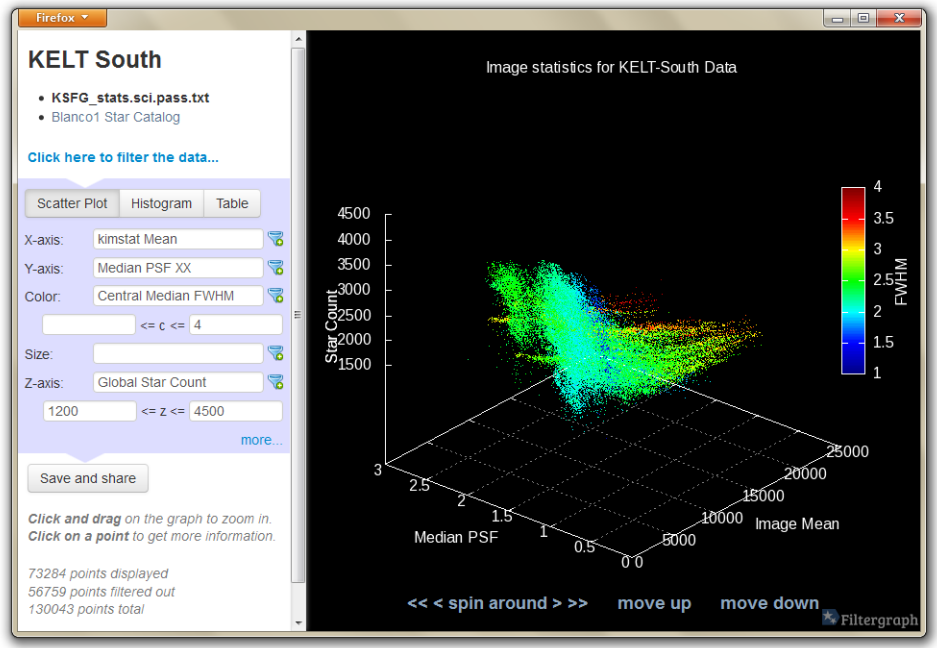

Figure 5. By applying the settings on the left to the publicly available Filtergraph portal at http://filtergraph.vanderbilt.edu/keltsouth, the three-dimensional scatter plot on the right can be generated. The title and axis labels have been customized, and the point type was set to plot using small dots. A version of this portal with these settings is available at http://filtergraph.vanderbilt.edu/7643704.

4. Technical details of Filtergraph

Filtergraph was written in Python 2.6.6 [17] and developed using the Web2py framework. Web2py is a full-stack web framework that allows web applications to be written in Python and deployed easily. Web2py was chosen for a variety of reasons. It can be deployed easily on any Windows, Macintosh or Linux machine, and is compatible with Apache for public access to the web server. It also comes with a secure and comprehensive administrative interface that can be used to edit the code, upload files, examine database entries, and view errors that have occurred [18].

A number of plugins and applications are used to support Filtergraph. On the server side, the Numpy package for Python is used for data storage, selection and manipulation [19]. The server also invokes the Gnuplot application for producing graphs [20] and the GraphicsMagick application for performing additional image manipulation [21]. On the client side, the ImgAreaSelect library is used to allow users to zoom in on the graph [22]. The JQuery [23], JQueryUI [24] and Bootstrap [25] libraries are also used to enhance the browser experience. Other third-party Python libraries used for processing data are ATpy [26], PyFITS [27], VO [28], and XLRD [29].

Filtergraph stores general information about each dataset in a MySQL database; the datasets themselves are stored on disk along with cache files. There are six tables in the database defined by Filtergraph, each of which store information about portals, datasets, administrators for each portal, headers contained in each dataset, and user feedback. There are also a few tables that are automatically generated by Web2py which stores user information. For security purposes, passwords are encrypted in the system using the SHA-512 algorithm.

An important need for Filtergraph is to generate images quickly for large datasets of up to millions of rows. To optimize it for speed, Filtergraph uses an "embarrassingly parallel system" to process the graph based on the MapReduce paradigm. [30] Once the data are loaded and processed using Numpy, the instructions and binary data are distributed equally among one to N instances of Gnuplot, where N is the number of cores in the system (eight at the time of this writing). This limit is imposed because running more than N processes at a time would not create a time advantage. Each instance of Gnuplot generates a PNG image containing its share of the data. The images are then merged together using Graphicsmagick and converted to the desired file format. The server returns the resulting image as well as information needed for the browser to support the zoom feature.

As the number of Gnuplot instances increases, it takes less time to generate the intermediate images but it takes more time to merge these images together. Filtergraph determines how many instances of Gnuplot to run by using equations to calculate how long it would take to generate the graph under one instance, under two instances and so on. The number of instances that would take the least time to generate the graph is then used. The equations are obtained using the Eureqa analytical software package on the basis of the times of graphs produced by Filtergraph under varying conditions. [31] [9] These equations are specific to our hardware setup, however in principle it should be possible to implement into Filtergraph the ability to determine these equations for any server setup, such as through a "benchmark" function to be added to the admin interface.

Since it is less practical to apply these procedures on Postscript and PDF files, generating these graphs always uses one process of Gnuplot. The same is true for histograms, since they typically require less time to generate. For obtaining tables and point information, Gnuplot and Graphicsmagick are bypassed entirely and the resulting data are returned directly to the browser.

## 4.1 Administration

Filtergraph is a proprietary web service which is developed and maintained by a programmer at Vanderbilt (currently the first author of this paper). An advisory board (currently including the co-authors of this paper) meets regularly to go through a list of suggested features in order to determine which should get priority. These features come from user feedback as well as discussions with Filtergraph users. New features are then added to the service as they are built and tested.

Additionally, Web2py produces a record each time it crashes. Once a crash occurs, users can optionally leave a report describing the problem that occurred. These records can be sorted by type later on so that the most frequently occurring bugs can be addressed.

## 5. Future Plans

Filtergraph currently has over 100 users in over 20 countries. We would like to see its use expanded beyond astronomy to other academic and non-academic fields where data are being heavily used. We also plan to add more features to Filtergraph such as improved statistical capabilities, additional color palettes and support for the Simple Application Messaging Protocol (SAMP).

Acknowledgements

The authors would like to thank Rachel-Chloe Gibbs for their help and support. In addition, the authors would like to acknowledge the users of Filtergraph who have provided valuable feedback throughout the development of this project, and support through NASA ADAP grant #NNX12AE22G and from the Vanderbilt Initiative in Data-intensive Astrophysics (VIDA): http://www.vanderbilt.edu/astro/vida/.